\documentclass[reqno,a4paper]{amsart}

\usepackage{amsmath,amssymb}             
\numberwithin{equation}{section}
\usepackage{amsfonts}
\usepackage{graphicx}
\usepackage{subfigure}
\usepackage{booktabs}
\usepackage{hyperref}
\hypersetup{
    colorlinks,%
    citecolor=blue,%
    filecolor=blue,%
    linkcolor=blue,%
    urlcolor=blue
}
\usepackage[authoryear]{natbib}
\usepackage{tikz}
\usetikzlibrary{patterns}
\usepackage{lineno}

\makeatletter
\g@addto@macro{\endabstract}{\@setabstract}
\newcommand{\authorfootnotes}{\renewcommand\thefootnote{\@fnsymbol\c@footnote}}%

\newtheorem{proposition}{Proposition}[section]

\newtheorem{theorem}[proposition]{Theorem}

\newtheorem{remark}[proposition]{Remark}

\newcommand{\selabel}[1]{\label{se:#1}}
\newcommand{\seref}[1]{Section~\ref{se:#1}}

\newcommand{\relabel}[1]{\label{re:#1}}

\def\l{{\lambda}}

\def\a{{\alpha}}

\def\G{{\Gamma}}

\def\s{{\sigma}}

\def\1{{\mathbf 1}}
\def\RR{{\mathbb R}}
\def\PP{{\mathbb P}}
\def\EE{{\mathbb E}}

\def\MR{{\mathcal R}}
\def\MM{{\mathcal M}}
\def\MN{{\mathcal N}}
\def\MB{{\mathcal B}}

\begin{document}
\bibliographystyle{plainnat}

\begin{center}
  \LARGE
{\bf Approximations for two-dimensional discrete scan statistics in some
block-factor type dependent models}\par \bigskip
  \normalsize
  Alexandru Am\u arioarei\textsuperscript{1,2,3}, Cristian Preda\textsuperscript{1,2}\par \bigskip

  \textsuperscript{1}Laboratoire de Math\'emathiques Paul Painlev\'e, UMR 8524, Univerit\'e de Sciences et Technologies de Lille 1, France \par
  \textsuperscript{2}INRIA Nord Europe/Modal, France\par
  \textsuperscript{3}National Institute of R\&D for Biological Sciences, Bucharest, Romania\par \bigskip

  \today
\end{center}

\email{alexandru.amarioarei@inria.fr}
\email{cristian.preda@polytech-lille.fr}

\begin{abstract}
{We consider the two-dimensional discrete scan statistic generated by a block-factor type model obtained from i.i.d. sequence. We present an approximation for the distribution of the scan statistics and the corresponding error bounds. A simulation study illustrates our methodology.}\@setabstract
\end{abstract}


\section{Introduction}\selabel{sec1}
\noindent
Let $N_{1}$, $N_{2}$ be positive integers, $\MR=[0,N_1]\times[0,N_2]$ be a rectangular region and $\left\{X_{i,j}  \ | \  1\leq i\leq N_{1}, 1\leq j\leq N_{2}\right\}$ be a family of random variables from a specified distribution. When the random variables $X_{i,j}$ take nonnegative integer values it is common to interpret them as the number of occurrences of some events observed in the elementary square sub-region $r_{i,j}=[i-1,i]\times [j-1,j]$. Let $m_1$, $m_{2}$ be positive integers such that $1\leq m_{1}\leq N_{1}$, $1\leq m_{2}\leq N_{2}$. For $1\leq i_1 \leq N_{1}-m_{1}+1$, $1\leq i_2\leq N_{2}-m_{2}+1$ define
\begin{equation}
Y_{i_1,i_2} =Y_{i_1,i_2}(m_{1},m_{2}) =\sum_{i=i_1}^{i_1+m_{1}-1}\sum_{j=i_2}^{i_2+m_{2}-1}X_{i,j}
\end{equation}
as the number of events in the rectangular region $\MR(i_1,i_2)=[i_1-1,i_1+m_1-1]\times[i_2-1,i_2+m_2-1]$, comprised of $m_1\times m_2$ adjacent elementary squares $r_{i,j}$. The {\em two-dimensional discrete scan statistic} is defined as the largest number of events in any rectangular scanning window $\MR(i_1,i_2)$, within the rectangular region $\MR$, i.e.
\begin{equation}\label{eq1}
S=S_{m_{1},m_{2}}(N_{1}, N_{2}) =\max_{\substack{1\leq i_1\leq N_1-m_1+1\\1\leq i_2\leq N_2-m_2+1}}{Y_{i_1,i_2}}.
\end{equation}
\par
\noindent
Most of research devoted to the two-dimensional discrete scan statistic considers the i.i.d. model for the random variables $X_{i,j}$. Then, the statistic $S$ is used for testing the null hypothesis of randomness ($H_0$), that assumes that $X_{i,j}$'s are independent and identically distributed according to some specified probability law, in general Bernoulli, binomial or Poisson (\citet{Glaz1996}, \citet{Glaz2001}), against an alternative ($H_1$) of clustering. Under $H_1$, one suppose that there is a change, with respect to $H_0$, in the distribution of the random field within a rectangular sub-region $\MR(i^*,j^*)\subset\MR$, with $1\leq i^* \leq N_{1}-m_{1}+1$, $1\leq j^*\leq N_{2}-m_{2}+1$, while outside this region $X_{i,j}$'s are distributed according to the null hypothesis distribution. As an example, consider that under $H_0$, $X_{i,j}$'s are i.i.d. Poisson random variables with mean $\l_0$. In this setting, the alternative hypothesis assumes that there exists a rectangular sub-region $\MR(i^*,j^*)$ such that for $i^*\leq i\leq i^*+m_1-1$ and $j^*\leq j\leq j^*+m_2-1$ the distribution of $X_{i,j}$ is given by a Poisson distribution of mean $\l_1>\l_0$ whereas in $\MR\setminus\MR(i^*,j^*)$ the events occur according to the distribution specified by the null hypothesis.
\par
\noindent
The distribution of the two-dimensional scan statistic,
\begin{equation*}
\displaystyle\PP\left(S_{m_1,m_2}(N_1,N_2)\leq n\right)
\end{equation*}
is successfully applied in brain imaging (\citet{Naiman2001}), astronomy (\citet{Darling1986}, \citet{Marcos2008}), target detection in sensors networks (\citet{Goerriero2009}), reliability theory (\citet{Boutsikas2000}) among many other domains. For an overview of the potential applications of the scan statistics one can refer to the monographs of \citet{Glaz2001} and more recently the one of \citet[Chapter 6]{Glaz2009}.
\par
\noindent
Since there are no exact formulas for $\PP(S\leq n)$, various methods of approximation and bounds have been proposed by several authors. An overview of these methods as well as a complete bibliography on the subject can be found in \citet{Glaz1996}, \citet[Chapter 16]{Glaz2001}, \citet{Bout2003}, \citet{HaimanPreda2} and the references therein.
\par
\noindent
In this paper we introduce a dependence structure for the underlying random field ($\left\{X_{i,j}  \ | \  1\leq i\leq N_{1}, 1\leq j\leq N_{2}\right\}$) based on a block-factor model and approximate the distribution of the two dimensional discrete scan statistics in this setting. Writing the scan statistics random variable $S$ as the maximum of a $1$-dependent stationary sequence, we approximate its distribution employing a result obtained by \citet{Haiman} and later improved by \citet{Amarioarei}. This approach was successfully used to evaluate the distribution of scan statistics, both in discrete and continuous cases, in a series of articles: for one-dimensional case in \citet{Haiman2} and \citet{Haiman2007}, for two-dimensional case in \citet{HaimanPreda2} and \citet{HaimanPreda} and for three-dimensional case in \citet{Amarioarei2013}. The advantage of our approach is that it can be applied under very general conditions and provides accurate approximations and sharp bounds for the errors.
\par\noindent
The paper is organized as follows. In \seref{sec2}, we introduce the block-factor type model that will generate the random field to be scanned. The methodology for approximating the
distribution of the scan statistics generated by the block-factor model as well as the associated error bounds are presented in \seref{sec3}. \seref{sec4} includes numerical results based on simulations for a particular block-factor model.
\section{Block-factor type model}\selabel{sec2}
\noindent
In this section we introduce a particular dependence structure for the random field $\left\{X_{i,j}  \ | \  1\leq i\leq N_{1}, 1\leq j\leq N_{2}\right\}$ based on a block-factor type model.
\par\noindent
Recall that (see \citet{Burton1993}) a sequence $(W_l)_{l\geq1}$ of random variables with state space $S_{W}$ is said to be a $k$ block-factor of the sequence $(\tilde{W}_l)_{l\geq1}$ with state space $S_{\tilde{W}}$, if there is a measurable function $f:S_{\tilde{W}}^k\to S_{W}$ such that
\begin{equation*}
W_l=f\left(\tilde{W}_l,\tilde{W}_{l+1},\dots,\tilde{W}_{l+k-1}\right)
\end{equation*}
for all $l$.
\par\noindent
Our block-factor type model is defined in the following way. Let $\tilde{N}_{1}$, $\tilde{N}_{2}$ be positive integers and $\left\{\tilde{X}_{i,j}  \ | \  1\leq i\leq \tilde{N}_{1}, 1\leq j\leq \tilde{N}_{2}\right\}$ be a family of independent and identically distributed real valued random variables. Notice that if the region $\tilde{\MR}=[0,\tilde{N}_1]\times[0,\tilde{N}_2]$ is divided in a grid with step $1$, then we can locate the random variables $\tilde{X}_{i,j}$ as being at the intersection of the $j$-th row with the $i$-th column.
\par\noindent
Let $x_1$, $x_2$, $y_1$, $y_2$ be nonnegative integers such that $x_1+x_2\leq \tilde{N}_{1}-1$ and $y_1+y_2\leq \tilde{N}_{2}-1$. Define $c_1 = x_1+x_2+1$, $c_2 = y_1+y_2+1$ and take $N_s=\tilde{N}_{s}-c_s+1$ for $s\in\{1,2\}$. To each pair $(i,j)\in\{x_1+1,\dots,\tilde{N}_1-x_2\}\times\{y_1+1,\dots,\tilde{N}_2-y_2\}$ we associate the random matrix of size $c_2\times c_1$, $C_{(i,j)}\in\MM_{c_2,c_1}(\RR)$, with entries
\begin{equation}\label{eq2}
C_{(i,j)}(k,l)=\tilde{X}_{i-x_1-1+l,j+y_2+1-k},\,\, \ 1\leq k\leq c_2,1\leq l\leq c_1.
\end{equation}
\par
\begin{figure}[ht]
\centering
\subfigure[]{%
\includegraphics[width=0.45\textwidth]{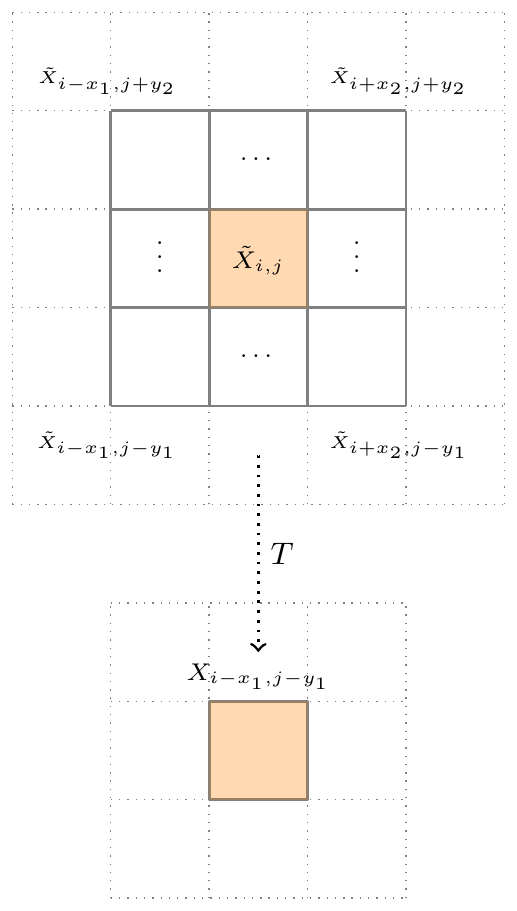}
\label{fig:subfigure1}}
\quad
\subfigure[]{%
\includegraphics[width=0.45\textwidth]{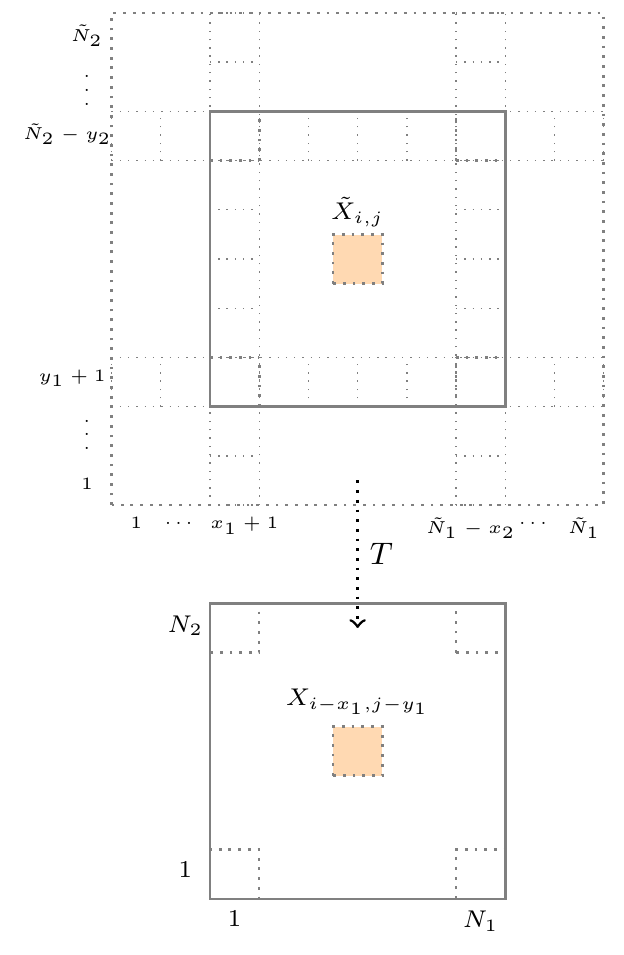}
\label{fig:subfigure2}}
\caption{Illustration of the block-factor type model}
\label{fig1}
\end{figure}

\noindent
If $T:\MM_{c_2,c_1}(\RR)\to\RR$ is a measurable function then the \textit{block-factor type} model is given by
\begin{equation}\label{eq3}
X_{i,j}=T\left(C_{(i+x_1,j+y_1)}\right)\, \text{with } 1\leq i\leq N_{1},\, 1\leq j\leq N_{2}.
\end{equation}
Figure~\ref{fig1} illustrates the construction of the block-factor model: on the left (see Fig~\ref{fig:subfigure1}) is presented the configuration matrix defined by Eq.\eqref{eq2} and the resulted random variable after applying the transformation $T$; on the right (see Fig~\ref{fig:subfigure2}) is exemplified how the i.i.d. model is transformed into the block-factor model .

\begin{figure}[ht]
 \centerline{
  \includegraphics[width=0.6\textwidth]{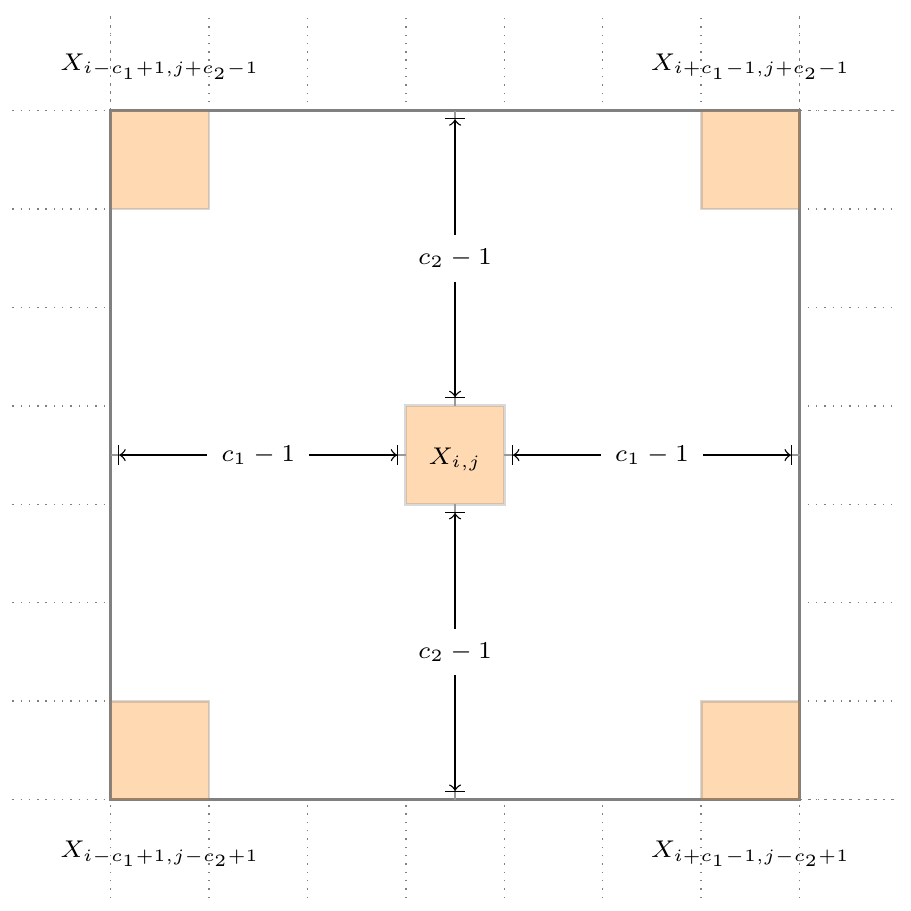}}
\caption{The dependence of $X_{i,j}$}
  \label{fig2}
\end{figure}

\noindent
Obviously, $\left\{X_{i,j}  \ | \  1\leq i\leq N_{1}, 1\leq j\leq N_{2}\right\}$ forms a dependent family of random variables (see Fig~\ref{fig2}).
\par\noindent
Recall that a sequence $(W_k)_{k\geq1}$ is $m$-dependent with $m\geq1$ (see \citet{Burton1993}), if for any $h\geq1$ the $\sigma$-fields generated by $\{W_1,\dots,W_h\}$ and $\{W_{h+m+1},\dots\}$ are independent. From the definition of the random variables $X_{i,j}$ given by Eq.\eqref{eq3}, we observe that for each $1\leq i\leq N_{1}$ the sequence $\left(X_{i,j}\right)_{1\leq j\leq N_{2}}$ is $(c_2-1)$-dependent and for each $1\leq j\leq N_{2}$ the sequence $\left(X_{i,j}\right)_{1\leq i\leq N_{1}}$ is $(c_1-1)$-dependent (see also Fig~\ref{fig2}).

\begin{remark}\relabel{rem2.2}
Notice that if $c_1=c_2=1$ ($x_1=x_2=0$ and $y_1=y_2=0$) then the sequence $X_{i,j}=\tilde{X}_{i,j}$ and we are in the i.i.d. situation.  In this case the distribution of the two-dimensional scan statistics can be approximated using the known methods (see \citet[Chapter 16]{Glaz2001} and \citet{Glaz2009}).
\par\noindent
If we take $\tilde{N}_2=1$, which automatically implies that $c_2=1$, we obtain an one dimensional block-factor model $W_i=X_{i,1}$ and the two-dimensional scan statistic becomes the usual discrete scan statistics in one dimension over a $(c_1-1)$-dependent sequence. The distribution of one dimensional scan statistics over this type of dependence was studied by \citet{HaimanPreda3} in the particular case of Gaussian stationary $1$-dependent ($x_1=0$, $x_2=1$ and $c_1=2$) sequences $W_i\sim\MN(0,1)$ of random variables generated by a two block-factor of the form
\begin{equation*}
W_i=aU_i+bU_{i+1},\ \ i\geq1,
\end{equation*}
where $a^2+b^2=1$ and $\left(U_i\right)_{i\geq1}$ is an i.i.d. sequence of $\MN(0,1)$ random variables.
\par\noindent
An application of the one dimensional scan statistics over a sequence of moving average of order $q$ ($c_1=q+1$) is presented in \seref{subsec4.2}.
\end{remark}
\par\noindent
Based on the model presented in this section, in \seref{sec3} we give an approximation for the distribution of two-dimensional scan statistic over the random field generated by the family $X_{i,j}$ and the corresponding error bounds.

\section{Approximation and error bounds}\selabel{sec3}
\noindent
In this section we present the methodology used to obtain the approximation of the two-dimensional discrete scan statistics distribution over the field generated by the block-factor model described in \seref{sec2}. Let's consider the scanning window of size $m_1\times m_2$ with $m_1\geq2$, $m_2\geq2$ and assume that for $s\in\{1,2\}$, $\tilde{N}_s=(L_s+1)(m_s+c_s-2)$ where $L_1$, $L_2$ are positive integers. Observe that
\begin{equation*}
N_s=L_s(m_s+c_s-2)+m_s-1,\, s\in\{1,2\}
\end{equation*}
and define the sequence
\begin{equation}\label{eq3.1}
Z_{k}=\max_{\substack{1\leq i_1\leq L_1(m_1+c_1-2)\\(k-1)(m_2+c_2-2)+1\leq i_2\leq k(m_2+c_2-2)}}{Y_{i_1,i_2}},\, \,\,\, k\in\{1,2,\dots,L_2\}.
\end{equation}
The random variables $Z_{k}$ represent the scan statistics on the overlapping $N_1\times 2(m_2+c_2-2)-(c_2-1)$ rectangular regions
\begin{equation*}
\MR_{k} = [1,N_1] \times [(k-1)(m_2+c_2-2)+1,(k+1)(m_2+c_2-2)-(c_2-1)].
\end{equation*}

\begin{figure}[ht]
  \centerline{
  \includegraphics[width=1\textwidth]{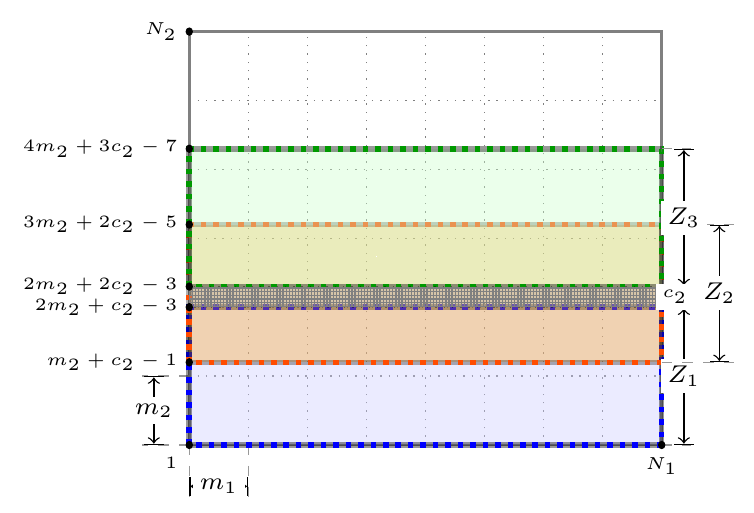}}
  \caption{Illustration of $Z_k$ emphasizing the $1$-dependence}
  \label{fig3}
\end{figure}

\begin{remark}\relabel{rem3.1}
If we consider the extreme situation when $c_2=m_2=1$ (or $c_1=m_1=1$), that is when we have row (column) independence and we are scanning only on rows (columns), then the sequence described by Eq.\eqref{eq3.1} is no longer well defined. In this case we define
\begin{equation}\label{eq3.2}
Z_{k}=\max_{1\leq i_1\leq L_1(m_1+c_1-2)}{Y_{i_1,k}},\, \,\,\, k\in\{1,2,\dots,L_2\},
\end{equation}
where $Y_{i_1,k}=\sum_{i=i_1}^{i_1+m_1-1}{X_{i,k}}.$
\end{remark}

\noindent
We observe that from Eq.\eqref{eq3.1} the set of random variables $\{Z_1,\dots,Z_{L_2}\}$ is 1-dependent (see also Figure~\ref{fig3}). Indeed, we have
\begin{align*}
Z_{k-1}&\in\sigma\left\{X_{i,j} | 1\leq i\leq N_1, (k-2)(m_2+c_2-2)+1\leq j\leq k(m_2+c_2-2)-(c_2-1)\right\}\\
      &\in\sigma\left\{\tilde{X}_{i,j} | 1\leq i\leq \tilde{N}_1, (k-2)(m_2+c-2-2)+1\leq j\leq k(m_2+c_2-2)\right\}
\end{align*}
and similarly,
\begin{align*}
Z_{k}&\in\sigma\left\{\tilde{X}_{i,j} | 1\leq i\leq \tilde{N}_1, (k-1)(m_2+c_2-2)+1\leq j\leq (k+1)(m_2+c_2-2)\right\},\\
Z_{k+1}&\in\sigma\left\{\tilde{X}_{i,j} | 1\leq i\leq \tilde{N}_1, k(m_2+c_2-2)+1\leq j\leq (k+2)(m_2+c_2-2)\right\}.
\end{align*}
From the above relations, the measurability of $T$ from the definition of the dependent model $\left\{X_{i,j}|1\leq i\leq N_1,1\leq j\leq N_2\right\}$ and the independence of the sequence $\left\{\tilde{X}_{i,j}|1\leq i\leq \tilde{N}_1,1\leq j\leq \tilde{N}_2\right\}$ we conclude that the sequence $\left(Z_k\right)_{1\leq k\leq L_2}$ is 1-dependent. Since $\tilde{X}_{i,j}$ are identically distributed we deduce stationarity of the random variables $Z_k$.
\par\noindent
Notice that from Eq.\eqref{eq3.1} and the definition of the two-dimensional scan statistics in Eq.\eqref{eq2} we have the following relation
\begin{equation}\label{eq3.3}
S = \max_{1\leq k\leq L_2}Z_{k}.
\end{equation}
The relation described by Eq.\eqref{eq3.3} is the key idea behind our approximation, i.e. the scan statistic random variable can be expressed as a maximum of 1-dependent stationary sequence of random variables. The approximation methodology that we use is based on the following result developed in \citet[Theorem 4]{Haiman} and improved in \citet[Theorem 2.6]{Amarioarei}:\\
Let $(W_k)_{k\geq1}$ be a stationary 1-dependent sequence of random variables and for $x<\sup\{u|\PP(W_1\leq u)<1\}$, consider
\begin{equation}\label{eq3.4}
q_m=q_m(x)=\PP(\max(W_1,\dots,W_m)\leq x).
\end{equation}

\begin{theorem}\label{T1}
Assume that $x$ is such that $q_1(x)\geq1-\a\geq 0.9$ and define $\eta=1+l\a$ with $l=l(\a)>t_2^3(\a)$ and $t_2(\a)$ the second root in magnitude of the equation $\a t^3-t+1=0$. Then the following relation holds
\begin{equation}\label{eqT1}
\left|q_m-\frac{2q_1-q_2}{\left[1+q_1-q_2+2(q_1-q_2)^2\right]^m}\right|\leq mF(\a,m)(1-q_1)^2,
\end{equation}
with
\begin{equation}\label{eqT1.2}
F(\a,m)=1+\frac{3}{m}+\left[\frac{\G(\a)}{m}+K(\a)\right](1-q_1),
\end{equation}
and where $\G(\a)=L(\a)+E(\a)$ and
\begin{align}
K(\a)&=\frac{\frac{11-3\a}{(1-\a)^2}+2l(1+3\a)\frac{2+3l\a-\a(2-l\a)(1+l\a)^2}{\left[1-\a(1+l\a)^2\right]^3}}{1-\frac{2\a(1+l\a)}{\left[1-\a(1+l\a)^2\right]^2}},\label{eqK1}\\
L(\a)&=3K(\a)(1+\a+3\a^2)[1+\a+3\a^2+K(\a)\a^3]+\a^6K^3(\a),\nonumber\\
     & +9\a(4+3\a+3\a^2)+55.1 \label{eqL1}\\
E(\a)&=\frac{\eta^5\left[1+(1-2\a)\eta\right]^4\left[1+\a(\eta-2)\right]\left[1+\eta+(1-3\a)\eta^2\right]}{2(1-\a\eta^2)^4\left[(1-\a\eta^2)^2-\a\eta^2(1+\eta-2\a\eta)^2\right]}.\label{eqE1}
\end{align}
\end{theorem}
\noindent
Following the approach in \citet{Amarioarei2013} for three dimensional scan statistics, we obtain an approximation formula for the distribution of two-dimensional scan statistic $S$ along with the corresponding error bounds, in two steps as follows.
\par\noindent
Define for $r\in\{2,3\}$,
\begin{equation}\label{eq3.5}
Q_r=Q_r(n)=\displaystyle\PP\left(\bigcap_{k=1}^{r-1}\{Z_k\leq n\}\right)=\displaystyle\PP\left(\max_{\substack{1\leq i_1\leq L_1(m_1+c_1-2)\\1\leq i_2\leq(r-1)(m_2+c_2-2)}}{Y_{i_1,i_2}}\leq n\right).
\end{equation}
For $n$ such that $Q_2(n)\geq1-\a_1\geq 0.9$, we apply the result in Theorem~\ref{T1} to obtain the first step approximation
\begin{equation}\label{eq3.6}
\left|\PP\left(S\leq n\right)-\frac{2Q_2-Q_3}{\left[1+Q_2-Q_3+2(Q_2-Q_3)^2\right]^{L_2}}\right|\leq L_2 F(\a_1,L_2)(1-Q_2)^2.
\end{equation}
In order to evaluate the approximation in Eq.\eqref{eq3.6} one has to find approximations for the quantities $Q_2$ and $Q_3$. To achieve this, we apply again the result of Theorem~\ref{T1}. We define, as in Eq.\eqref{eq3.1}, for each $r\in\{2,3\}$ and $l\in\{1,2,\dots,L_1\}$ the random variables
\begin{equation}\label{eq3.7}
Z^{(r)}_{l}=\max_{\substack{(l-1)(m_1+c_1-2)+1\leq i_1\leq l(m_1+c_1-2)\\1\leq i_2\leq (r-1)(m_2+c_2-2)}}{Y_{i_1,i_2}}.
\end{equation}
As described in the case of the sequence $Z_k$, we deduce that the random variables $Z^{(r)}_{l}$ defined by Eq.\eqref{eq3.7} are stationary, $1$-dependent and the following relation holds:
\begin{equation}\label{eq3.8}
Q_r=\PP\left(\max_{1\leq l\leq L_1}Z^{(r)}_{l}\leq n\right),\ \ \ r\in\{2,3\}.
\end{equation}
Denoting, for $u,v\in\{2,3\}$
\begin{equation}\label{eq3.9}
Q_{uv}=Q_{uv}(n)=\displaystyle\PP\left(\bigcap_{l=1}^{u-1}\{Z^{(v)}_l\leq n\}\right)=\displaystyle\PP\left(\max_{\substack{1\leq i_1\leq (u-1)(m_1+c_1-2)\\1\leq i_2\leq(v-1)(m_2+c_2-2)}}{Y_{i_1,i_2}}\leq n\right)
\end{equation}
then, under the supplementary condition that $n$ is such that $Q_{23}(n)\geq1-\a_2\geq 0.9$, we apply Theorem~\ref{T1} to obtain
\begin{equation}\label{eq3.10}
\left|Q_r-\frac{2Q_{2r}-Q_{3r}}{\left[1+Q_{2r}-Q_{3r}+2(Q_{2r}-Q_{3r})^2\right]^{L_1}}\right|\leq L_1 F(\a_2,L_1)(1-Q_{2r})^2.
\end{equation}
Combining Eq.\eqref{eq3.6} and Eq.\eqref{eq3.10} we find an approximation formula for the distribution of the two-dimensional scan statistic depending on the values of $Q_{22}$, $Q_{23}$, $Q_{32}$ and $Q_{33}$. There are no exact formulas for $Q_{uv}$, $u,v\in\{2,3\}$, thus these quantities will be evaluated using Monte Carlo simulation. The approximation process is summarized by the diagram in Figure~\ref{fig4}:

\begin{figure}[ht]
  \centerline{
  \includegraphics[width=1\textwidth]{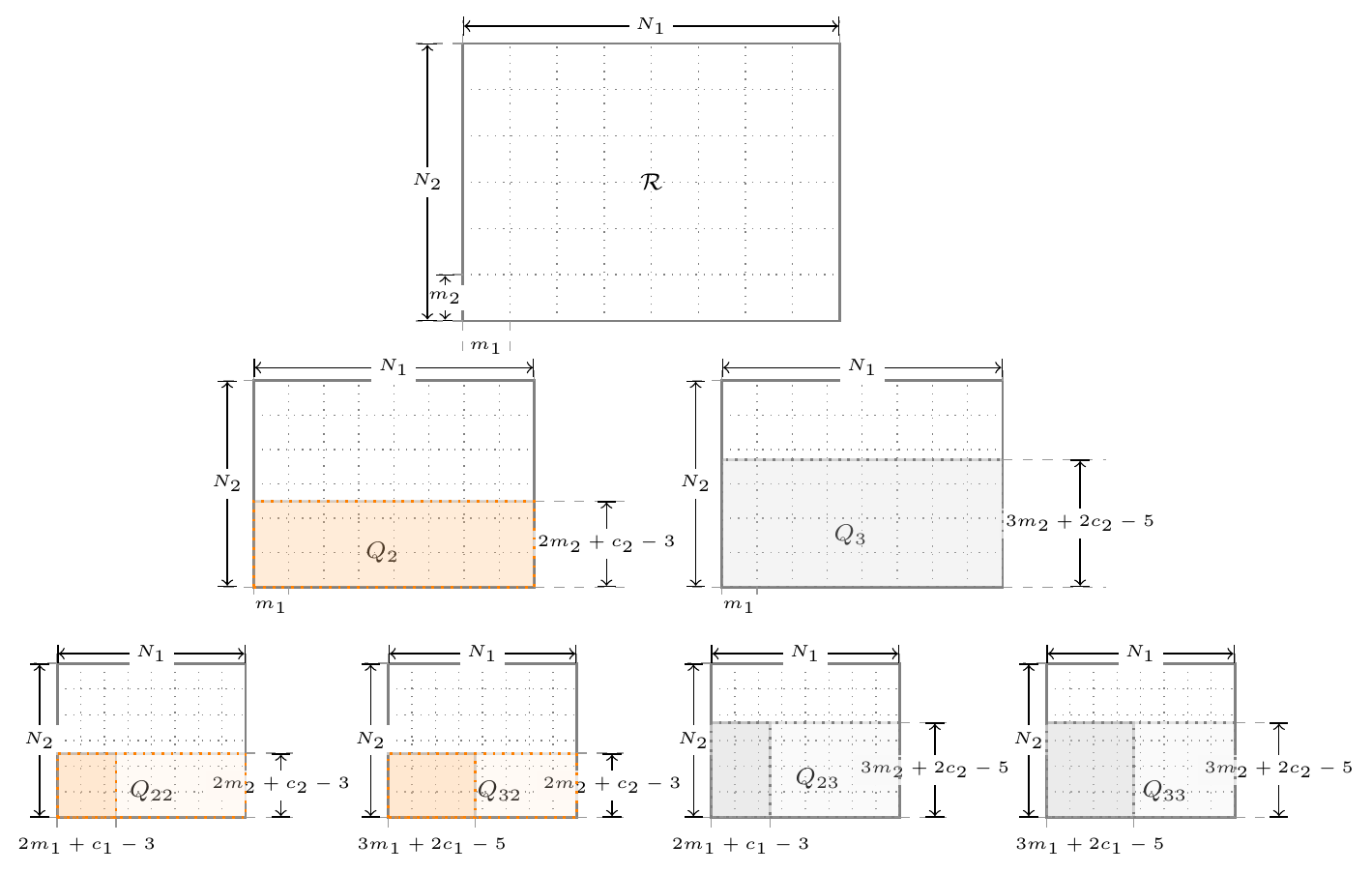}}
  \caption{Illustration of the approximation process}
  \label{fig4}
\end{figure}

\begin{remark}\relabel{rem3.2}
If $\tilde{N}_1$ and $\tilde{N}_2$ are not multiples of $m_1+c_1-2$ and $m_2+c_2-2$, respectively, then we take $L_j+1=\left\lfloor{\frac{\tilde{N}_j}{m_j+c_j-2}}\right\rfloor$ for $j\in\{1,2\}$. Based on the inequalities
\begin{equation}\label{eq3.11}
\PP\left(S_{m_1,m_2}(M_1,M_2)\leq n\right)\leq\PP\left(S_{m_1,m_2}(N_1,N_2)\leq n\right)\leq\PP\left(S_{m_1,m_2}(T_1,T_2)\leq n\right),
\end{equation}
where for $j\in\{1,2\}$ we consider $M_j=(L_j+2)(m_j+c_j-2)-(c_j-1)$ and $T_j=(L_j+1)(m_j+c_j-2)-(c_j-1)$, we can approximate the distribution of the scan statistics by linear interpolation.
\end{remark}
\subsection{Computing the error bounds}\selabel{subsec31}
\noindent
For the error computation we have to notice that there are three expressions involved: the first one is the \textit{theoretical error} ($E_{app}$) obtained from the substitution of Eq.\eqref{eq3.10} in Eq.\eqref{eq3.6} whereas the other two are \textit{ simulations errors}, one corresponding to the approximation formula ($E_{sf}$) and the other to the error formula ($E_{sapp}$). In what follows we will deal with each of them separately. To simplify the presentation it will be convenient to introduce the following notations:
\begin{align}
H(x,y,m)&=\frac{2x-y}{[1+x-y+2(x-y)^2]^{m}},\ \a_1=1-Q_3,\ \a_{2}=1-Q_{23},\nonumber\\
F_1&=F(\a_2,L_1),\ F_2=F(\a_{1},L_2),\ R_s=H\left(Q_{2s},Q_{3s},L_1\right),\ s\in\{2,3\}.\nonumber
\end{align}
Notice that the choice for the thresholds $\a_1$ and $\a_2$ is natural since we have the inequalities $Q_3\leq Q_2$ and $Q_{23}\leq Q_{22}$. Based on mean value theorem in two dimensions, one can easily verify that if $y_i\leq x_i$, $i\in\{1,2\}$ then we have the relation
\begin{equation}\label{eq3.1.1}
\left|H(x_1,y_1,m)-H(x_2,y_2,m)\right|\leq m\left[|x_1-x_2|+|y_1-y_2|\right].
\end{equation}
Rewriting Eq.\eqref{eq3.6} using the above notations and applying the inequality in Eq.\eqref{eq3.1.1} we can write
\begin{align}
\left|\PP(S\leq n)-H\left(R_2,R_3,L_2\right)\right|&\leq\left|\PP(S\leq n)-H\left(Q_2,Q_3,L_2\right)\right|+\nonumber\\
                                              &\left|H\left(Q_2,Q_3,L_2\right)-H\left(R_2,R_3,L_2\right)\right|\nonumber\\
                                              &\leq L_2F_2\left(1-Q_2\right)^2+L_2\left[|Q_2-R_2|+|Q_3-R_3|\right].\label{eq3.1.2}
\end{align}
If we substitute Eq.\eqref{eq3.10} in Eq.\eqref{eq3.1.2} and take $B_2=1-R_2+L_1F_1(1-Q_{22})^2$, then the theoretical approximation error is given by
\begin{equation}\label{eq3.1.3}
E_{app}=L_2F_2B_2^2+L_1L_2F_1\left[(1-Q_{22})^2+(1-Q_{23})^2\right].
\end{equation}
To compute the simulation error corresponding to the approximation formula let us denote with $\hat{Q}_{uv}$ the simulated values corresponding to $Q_{uv}$ for each $u,v\in\{2,3\}$. Usually between the true and the estimated values we have a relation of the type
\begin{equation}\label{eq3.1.4}
\left|Q_{uv}-\hat{Q}_{uv}\right|\leq \beta_{uv}.
\end{equation}
Indeed, if $ITER$ is the number of iterations used in the Monte Carlo simulation algorithm for the estimation of $Q_{uv}$ then, one can consider, for example, the bound $\beta_{uv}=1.96\sqrt{\frac{\hat{Q}_{uv}(1-\hat{Q}_{uv})}{ITER}}$ with a $95\%$ confidence level. Taking for $r\in\{2,3\}$, $\hat{Q}_{r}=H\left(\hat{Q}_{2r},\hat{Q}_{3r},L_1\right)$ to be the simulated values that corresponds to $Q_{r}$ and applying Eq.\eqref{eq3.1.1} whenever $\hat{Q}_{3}\leq\hat{Q}_{2}$ we get
\begin{align}
\left|H\left(R_2,R_3,L_2\right)-H\left(\hat{Q}_{2},\hat{Q}_{3},L_2\right)\right|&\leq L_2\left[\left|R_2-\hat{Q}_{2}\right|+\left|R_3-\hat{Q}_{3}\right|\right]\nonumber\\
                        &\leq L_1L_2\left[\left|Q_{22}-\hat{Q}_{22}\right|+\left|Q_{23}-\hat{Q}_{23}\right|+\right.\nonumber\\
                        &\left.\left|Q_{32}-\hat{Q}_{32}\right|+\left|Q_{33}-\hat{Q}_{33}\right|\right].\label{eq3.1.5}
\end{align}
Combining Eq.\eqref{eq3.1.5} and Eq.\eqref{eq3.1.4} we obtain the simulation error associated with the approximation formula
\begin{equation}\label{eq3.1.6}
E_{sf}=L_1L_2(\beta_{22}+\beta_{23}+\beta_{32}+\beta_{33}).
\end{equation}
Finally, introducing
\begin{align*}
C_{2v}&=1-\hat{Q}_{2v}+\beta_{2v},\ \ v\in\{2,3\},\\
C_2&=1-\hat{Q}_{2}+L_1(\beta_{22}+\beta_{32})+L_1F_1C_{22}^2,
\end{align*}
and substituting them in the theoretical approximation error formula in Eq.\eqref{eq3.1.3}, we obtain the simulation error corresponding to the approximation error formula
\begin{equation}\label{eq3.1.7}
E_{sapp}=L_2F_2C_2^2+L_1L_2F_1\left[C_{22}^2+C_{23}^2\right].
\end{equation}
Adding the expressions from Eq.\eqref{eq3.1.3}, Eq.\eqref{eq3.1.6} and Eq.\eqref{eq3.1.7} we have the total error,
\begin{equation}\label{eq3.1.8}
E_{total}=E_{app}+E_{sf}+E_{sapp}.
\end{equation}
\section{Examples and numerical results}\selabel{sec4}
\noindent
In order to illustrate the efficiency of the approximation and the error bounds obtained in \seref{sec3}, we consider the following examples: a \textit{minesweeper game} presented in \seref{subsec4.1} and an one dimensional scan statistics over a moving average model described in \seref{subsec4.2}.
\subsection{Example 1: minesweeper game}\selabel{subsec4.1}
\noindent
Let $\tilde{N}_1$, $\tilde{N}_2$ be positive integers and $\left\{\tilde{X}_{i,j}  \ | \  1\leq i\leq \tilde{N}_{1}, 1\leq j\leq \tilde{N}_{2}\right\}$ be a family of i.i.d. Bernoulli random variables of parameter $p$. We interpret the random variable $\tilde{X}_{i,j}$ as representing the presence ($\tilde{X}_{i,j}=1$) or absence ($\tilde{X}_{i,j}=0$) of a mine in the elementary square region $\tilde{r}_{i,j}=[i-1,i]\times[j-1,j]$.
\par
\begin{figure}[ht]
 \centerline{
  \includegraphics[width=1\textwidth]{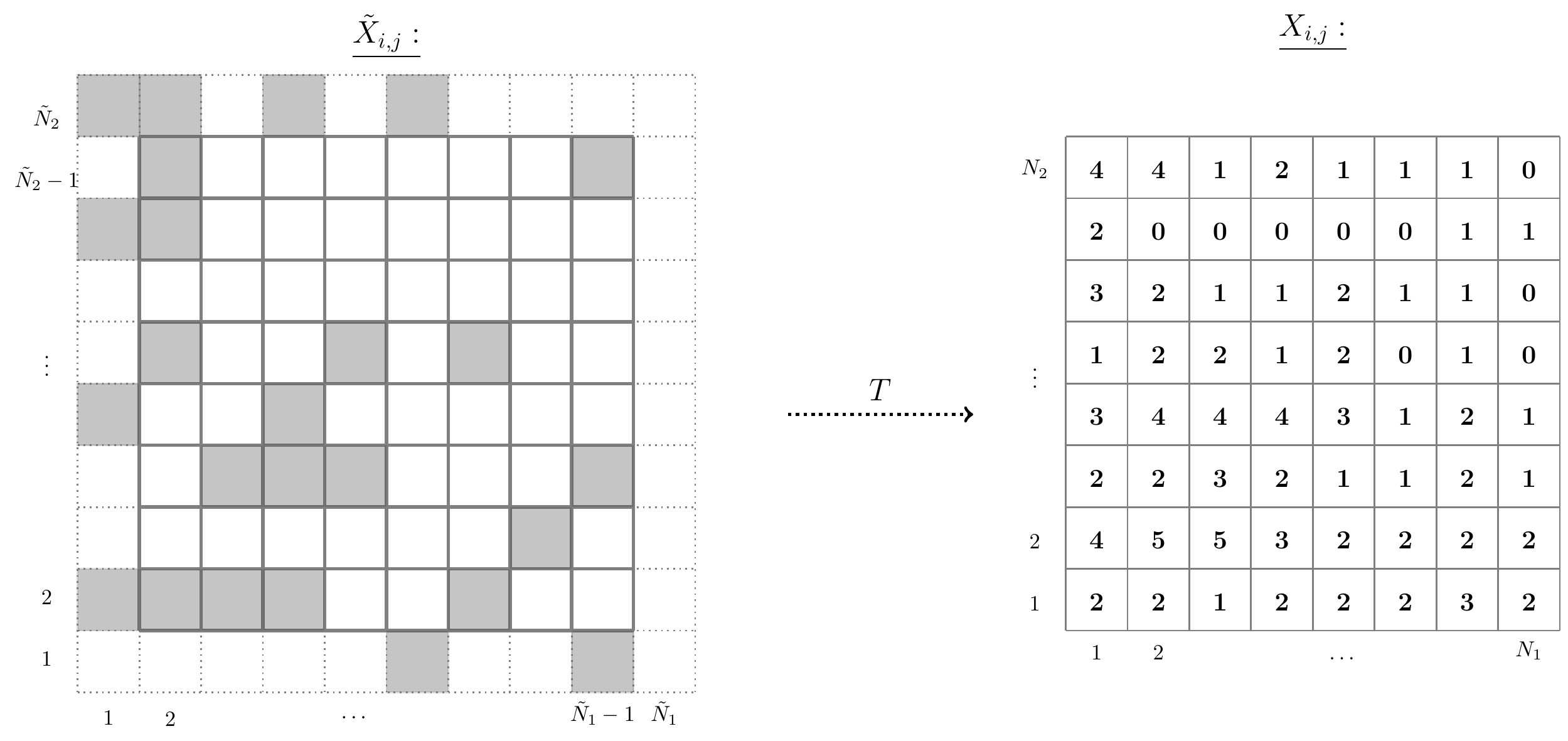}}
  \caption{A realization of the minesweeper related model}
  \label{fig5}
\end{figure}

\noindent
In this example we consider $x_1=x_2=1$ and $y_1=y_2=1$. Based on the notations introduced in \seref{sec2}, we observe that $c_1=c_2=3$, $N_1=\tilde{N}_1-2$ and $N_2=\tilde{N}_2-2$. For each $(i,j)\in\{2,\dots,\tilde{N}_1-1\}\times\{2,\dots,\tilde{N}_2-1\}$ the configuration matrix is given by
\begin{equation}\label{eq4.1}
C_{(i,j)}=\left(C_{(i,j)}(k,l)\right)_{\substack{1\leq k\leq 3\\1\leq l\leq 3}},\ \text{where}\ \ C_{(i,j)}(k,l)=\tilde{X}_{i+l-2,j+2-k}.
\end{equation}
Let $T:\MM_{3,3}(\RR)\to\RR$
\begin{equation}\label{eq4.2}
T\left(\begin{array}{ccc}
        a_{11} & a_{12} & a_{13}\\
        a_{21} & a_{22} & a_{23}\\
        a_{31} & a_{32} & a_{33}
\end{array}
\right)=\displaystyle\sum_{1\leq s,t \leq 3}{a_{st}-a_{22}}
\end{equation}
and define for $1\leq i\leq N_{1}$ and $1\leq j\leq N_{2}$, the block-factor model
\begin{equation}\label{eq4.3}
X_{i,j}=T\left(C_{(i+1,j+1)}\right)=\displaystyle\sum_{\substack{(s,t)\in\{0,1,2\}^2\\(s,t)\neq(1,1)}}{\tilde{X}_{i+s,j+t}}.
\end{equation}
The random variable $X_{i,j}$ can be interpreted as the number of neighboring mines associated with the location $(i,j)$. In Figure~\ref{fig5} we present a realization of the introduced model. On the left, we have the realization of the initial set of random variables where the gray squares represent the presence of mines while the white squares signifies the absence of mines. On the right side we have the realization of the $X_{i,j}$ random variables, that is the corresponding number of neighboring mines associated to each site.

\noindent
We present numerical results (Table~\ref{tab1a}-Table~\ref{tab4b}) for the described block-factor model with $\tilde{N}_1=\tilde{N}_2=44$ (that is $N_1=N_2=42$), $m_1=m_2=3$ and the underlying random field generated by i.i.d. Bernoulli random variables of parameter $p$ ($\tilde{X}_{i,j}\sim\MB(p)$) in the range $\{0.1,0.3,0.5,0.7\}$. We also include numerical values for the corresponding i.i.d. model: $N_1=N_2=42$, $m_1=m_2=3$ and $X_{i,j}\sim\MB(8,p)$.


\begin{table}[ht!]
    \caption{Block-factor: $m_1=m_2=3$, $\tilde{N}_1=\tilde{N}_2=44$, $N_1=N_2=42$, $\bf{ p=0.1}$, $ITER=10^8$}
\centering
\small{
 \begin{tabular}{cccccc}
\toprule
$n$ &$Sim$  & $Approx$  & $E_{app}$  & $E_{sim}$ & $E_{total}$  \\
\midrule
29 & 0.828763 & 0.813457 & 0.018678 & 0.024528 & 0.043205 \\
30 & 0.886702 & 0.875875 & 0.006135 & 0.010670 & 0.016805 \\
31 & 0.930094 & 0.922997 & 0.001912 & 0.005374 & 0.007286 \\
32 & 0.957297 & 0.953079 & 0.000628 & 0.003290 & 0.003918 \\
33 & 0.974541 & 0.971980 & 0.000204 & 0.002239 & 0.002443 \\
34 & 0.985523 & 0.984022 & 0.000063 & 0.001588 & 0.001651 \\
35 & 0.991524 & 0.990718 & 0.000020 & 0.001171 & 0.001191 \\
36 & 0.995301 & 0.994885 & 0.000006 & 0.000854 & 0.000860 \\
37 & 0.997492 & 0.997253 & 0.000002 & 0.000617 & 0.000619 \\
38 & 0.998668 & 0.998547 & 0.000000 & 0.000447 & 0.000447 \\
39 & 0.999313 & 0.999272 & 0.000000 & 0.000319 & 0.000319 \\
40 & 0.999653 & 0.999629 & 0.000000 & 0.000231 & 0.000231 \\
41 & 0.999826 & 0.999808 & 0.000000 & 0.000164 & 0.000164 \\
42 & 0.999916 & 0.999911 & 0.000000 & 0.000116 & 0.000116 \\
43 & 0.999963 & 0.999959 & 0.000000 & 0.000079 & 0.000079 \\
44 & 0.999981 & 0.999979 & 0.000000 & 0.000054 & 0.000054 \\
45 & 0.999991 & 0.999993 & 0.000000 & 0.000037 & 0.000037 \\
46 & 0.999995 & 0.999997 & 0.000000 & 0.000022 & 0.000022 \\
47 & 0.999999 & 0.999999 & 0.000000 & 0.000017 & 0.000017 \\
48 & 1.000000 & 0.999999 & 0.000000 & 0.000009 & 0.000009 \\

\bottomrule
\end{tabular}
}
\setlength{\tabcolsep}{20em}
\label{tab1a}
\end{table}

\begin{table}[ht!]
    \caption{Independent: $m_1=m_2=3$, $N_1=N_2=42$, $\bf{\MB(n=8, p=0.1)}$, $ITER=10^5$}
\centering
\small{
 \begin{tabular}{cccccc}
\toprule
$n$ &$Sim$  & $Approx$  & $E_{app}$  & $E_{sim}$ & $E_{total}$  \\
\midrule
17 & 0.789376 & 0.788934 & 0.005813 & 0.011393 & 0.017206  \\
18 & 0.925456 & 0.925186 & 0.000529 & 0.002095 & 0.002625  \\
19 & 0.976889 & 0.976763 & 0.000045 & 0.000455 & 0.000500  \\
20 & 0.993444 & 0.993447 & 0.000003 & 0.000105 & 0.000108  \\
21 & 0.998288 & 0.998287 & 0.000000 & 0.000023 & 0.000024  \\
22 & 0.999584 & 0.999583 & 0.000000 & 0.000005 & 0.000005  \\
23 & 0.999905 & 0.999905 & 0.000000 & 0.000001 & 0.000001  \\
24 & 0.999980 & 0.999980 & 0.000000 & 0.000000 & 0.000000  \\

\bottomrule
\end{tabular}
}
\setlength{\tabcolsep}{20em}
\label{tab1b}
\end{table}


\begin{table}[ht!]
    \caption{Block-factor: $m_1=m_2=3$, $\tilde{N}_1=\tilde{N}_2=44$, $N_1=N_2=42$, $\bf{ p=0.3}$, $ITER=10^8$}
\centering
\small{
 \begin{tabular}{cccccc}
\toprule
$n$ &$Sim$  & $Approx$  & $E_{app}$  & $E_{sim}$ & $E_{total}$  \\
\midrule
48 & 0.768889 & 0.749275 & 0.046577 & 0.053831 & 0.100408 \\
49 & 0.844717 & 0.829918 & 0.014207 & 0.019702 & 0.033908 \\
50 & 0.899398 & 0.889501 & 0.004574 & 0.008810 & 0.013384 \\
51 & 0.936771 & 0.930795 & 0.001499 & 0.004769 & 0.006269 \\
52 & 0.961836 & 0.958113 & 0.000485 & 0.002988 & 0.003472 \\
53 & 0.977672 & 0.975326 & 0.000152 & 0.002045 & 0.002197 \\
54 & 0.987307 & 0.985922 & 0.000047 & 0.001463 & 0.001510 \\
55 & 0.993022 & 0.992251 & 0.000014 & 0.001056 & 0.001070 \\
56 & 0.996333 & 0.995917 & 0.000004 & 0.000761 & 0.000765 \\
57 & 0.998151 & 0.997954 & 0.000001 & 0.000539 & 0.000540 \\
58 & 0.999091 & 0.998992 & 0.000000 & 0.000381 & 0.000381 \\
59 & 0.999576 & 0.999522 & 0.000000 & 0.000265 & 0.000265 \\
60 & 0.999794 & 0.999802 & 0.000000 & 0.000178 & 0.000178 \\
61 & 0.999908 & 0.999920 & 0.000000 & 0.000115 & 0.000115 \\
62 & 0.999965 & 0.999973 & 0.000000 & 0.000077 & 0.000077 \\
63 & 0.999993 & 0.999991 & 0.000000 & 0.000044 & 0.000044 \\
64 & 0.999999 & 0.999998 & 0.000000 & 0.000028 & 0.000028 \\
65 & 1.000000 & 0.999999 & 0.000000 & 0.000017 & 0.000017 \\
\bottomrule
\end{tabular}
}
\setlength{\tabcolsep}{20em}
\label{tab2a}
\end{table}

\begin{table}[ht!]
    \caption{Independent: $m_1=m_2=3$, $N_1=N_2=42$, $\bf{\MB(n=8, p=0.3)}$, $ITER=10^5$}
\centering
\small{
 \begin{tabular}{cccccc}
\toprule
$n$ &$Sim$  & $Approx$  & $E_{app}$  & $E_{sim}$ & $E_{total}$  \\
\midrule
35 & 0.716804 & 0.716395 & 0.012836 & 0.021243 & 0.034079  \\
36 & 0.867167 & 0.866643 & 0.001951 & 0.005093 & 0.007044  \\
37 & 0.943946 & 0.944024 & 0.000285 & 0.001409 & 0.001694  \\
38 & 0.978505 & 0.978400 & 0.000039 & 0.000419 & 0.000457  \\
39 & 0.992274 & 0.992262 & 0.000005 & 0.000126 & 0.000131  \\
40 & 0.997395 & 0.997399 & 0.000001 & 0.000037 & 0.000037  \\
41 & 0.999176 & 0.999178 & 0.000000 & 0.000010 & 0.000010  \\
42 & 0.999753 & 0.999754 & 0.000000 & 0.000003 & 0.000003  \\
43 & 0.999931 & 0.999931 & 0.000000 & 0.000001 & 0.000001  \\
44 & 0.999982 & 0.999982 & 0.000000 & 0.000000 & 0.000000  \\
45 & 0.999995 & 0.999995 & 0.000000 & 0.000000 & 0.000000  \\
\bottomrule
\end{tabular}
}
\setlength{\tabcolsep}{20em}
\label{tab2b}
\end{table}


\begin{table}[ht!]
    \caption{Block-factor: $m_1=m_2=3$, $\tilde{N}_1=\tilde{N}_2=44$, $N_1=N_2=42$, $\bf{ p=0.5}$, $ITER=10^8$}
\centering
\small{
 \begin{tabular}{cccccc}
\toprule
$n$ &$Sim$  & $Approx$  & $E_{app}$  & $E_{sim}$ & $E_{total}$  \\
\midrule
61 & 0.725109 & 0.701781 & 0.085110 & 0.093544 & 0.178654 \\
62 & 0.828019 & 0.888902 & 0.004453 & 0.008665 & 0.013118 \\
63 & 0.899560 & 0.888902 & 0.004453 & 0.008665 & 0.013118 \\
64 & 0.945304 & 0.939436 & 0.001049 & 0.004054 & 0.005103 \\
65 & 0.972203 & 0.969026 & 0.000235 & 0.002334 & 0.002569 \\
66 & 0.986999 & 0.985439 & 0.000047 & 0.001460 & 0.001507 \\
67 & 0.994506 & 0.993814 & 0.000008 & 0.000927 & 0.000935 \\
68 & 0.997851 & 0.997605 & 0.000001 & 0.000572 & 0.000573 \\
69 & 0.999326 & 0.999230 & 0.000000 & 0.000320 & 0.000320 \\
70 & 0.999826 & 0.999786 & 0.000000 & 0.000171 & 0.000171 \\
71 & 0.999968 & 0.999952 & 0.000000 & 0.000083 & 0.000083 \\
72 & 1.000000 & 1.000000 & 0.000000 & 0.000000 & 0.000000 \\
\bottomrule
\end{tabular}
}
\setlength{\tabcolsep}{20em}
\label{tab3a}
\end{table}

\begin{table}[ht!]
    \caption{Independent: $m_1=m_2=3$, $N_1=N_2=42$, $\bf{\MB(n=8, p=0.5)}$, $ITER=10^5$}
\centering
\small{
 \begin{tabular}{cccccc}
\toprule
$n$ &$Sim$  & $Approx$  & $E_{app}$  & $E_{sim}$ & $E_{total}$  \\
\midrule
50 & 0.741089 & 0.735210 & 0.010514 & 0.018002 & 0.028516  \\
51 & 0.882209 & 0.880827 & 0.001499 & 0.004196 & 0.005695  \\
52 & 0.952545 & 0.952389 & 0.000200 & 0.001098 & 0.001299  \\
53 & 0.982842 & 0.982891 & 0.000024 & 0.000307 & 0.000331  \\
54 & 0.994328 & 0.994337 & 0.000002 & 0.000084 & 0.000087  \\
55 & 0.998282 & 0.998278 & 0.000000 & 0.000022 & 0.000022  \\
56 & 0.999517 & 0.999518 & 0.000000 & 0.000005 & 0.000005  \\
57 & 0.999876 & 0.999876 & 0.000000 & 0.000001 & 0.000001  \\
58 & 0.999971 & 0.999971 & 0.000000 & 0.000000 & 0.000000  \\
59 & 0.999994 & 0.999994 & 0.000000 & 0.000000 & 0.000000  \\
60 & 0.999999 & 0.999999 & 0.000000 & 0.000000 & 0.000000  \\
\bottomrule
\end{tabular}
}
\setlength{\tabcolsep}{20em}
\label{tab3b}
\end{table}


\begin{table}[ht!]
    \caption{Block-factor: $m_1=m_2=3$, $\tilde{N}_1=\tilde{N}_2=44$, $N_1=N_2=42$, $\bf{ p=0.7}$, $ITER=10^8$}
\centering
\small{
 \begin{tabular}{cccccc}
\toprule
$n$ &$Sim$  & $Approx$  & $E_{app}$  & $E_{sim}$ & $E_{total}$  \\
\midrule
70 & 0.729239 & 0.705944 & 0.074290 & 0.082392 & 0.156682 \\
71 & 0.876484 & 0.864370 & 0.006976 & 0.011623 & 0.018600 \\
72 & 1.000000 & 1.000000 & 0.000000 & 0.000000 & 0.000000 \\
\bottomrule
\end{tabular}
}
\setlength{\tabcolsep}{20em}
\label{tab4a}
\end{table}

\begin{table}[ht!]
    \caption{Independent: $m_1=m_2=3$, $N_1=N_2=42$, $\bf{\MB(n=8, p=0.7)}$, $ITER=10^5$}
\centering
\small{
 \begin{tabular}{cccccc}
\toprule
$n$ &$Sim$  & $Approx$  & $E_{app}$  & $E_{sim}$ & $E_{total}$  \\
\midrule
62.0 & 0.620295 & 0.611819 & 0.030328 & 0.042319 & 0.072646  \\
63.0 & 0.847421 & 0.846730 & 0.002591 & 0.005851 & 0.008442  \\
64.0 & 0.952524 & 0.952588 & 0.000194 & 0.000978 & 0.001172  \\
65.0 & 0.987854 & 0.987887 & 0.000011 & 0.000168 & 0.000179  \\
66.0 & 0.997472 & 0.997460 & 0.000000 & 0.000026 & 0.000027  \\
67.0 & 0.999568 & 0.999568 & 0.000000 & 0.000003 & 0.000003  \\
68.0 & 0.999943 & 0.999943 & 0.000000 & 0.000000 & 0.000000  \\
69.0 & 0.999994 & 0.999994 & 0.000000 & 0.000000 & 0.000000  \\
\bottomrule
\end{tabular}
}
\setlength{\tabcolsep}{20em}
\label{tab4b}
\end{table}

\begin{figure}[ht!]
 \centerline{
  \includegraphics[width=0.65\textwidth]{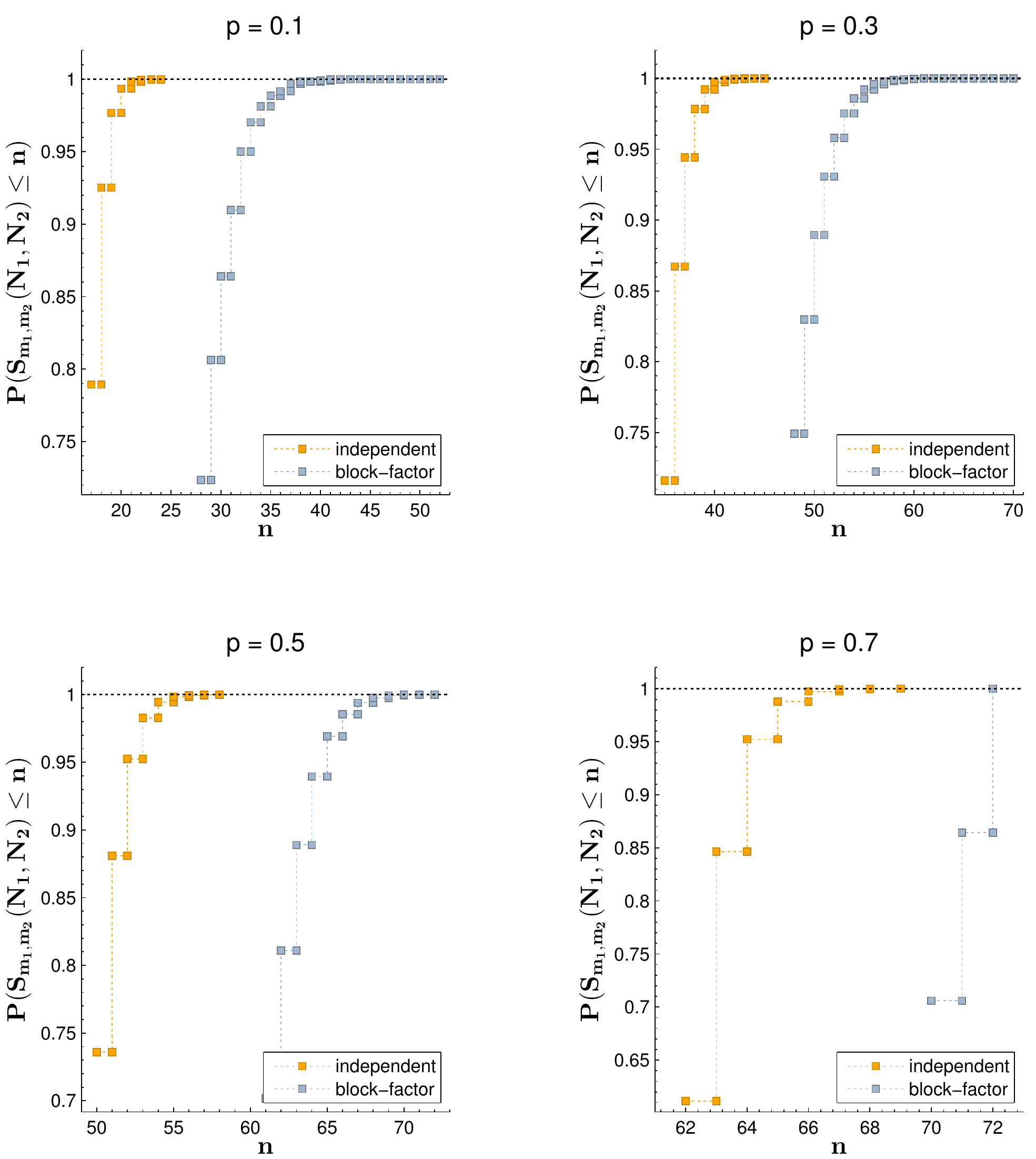}}
  \caption{Cumulative distribution function for block--factor and i.i.d. models}
  \label{fig6}
\end{figure}

\begin{figure}[ht!]
 \centerline{
  \includegraphics[width=0.65\textwidth]{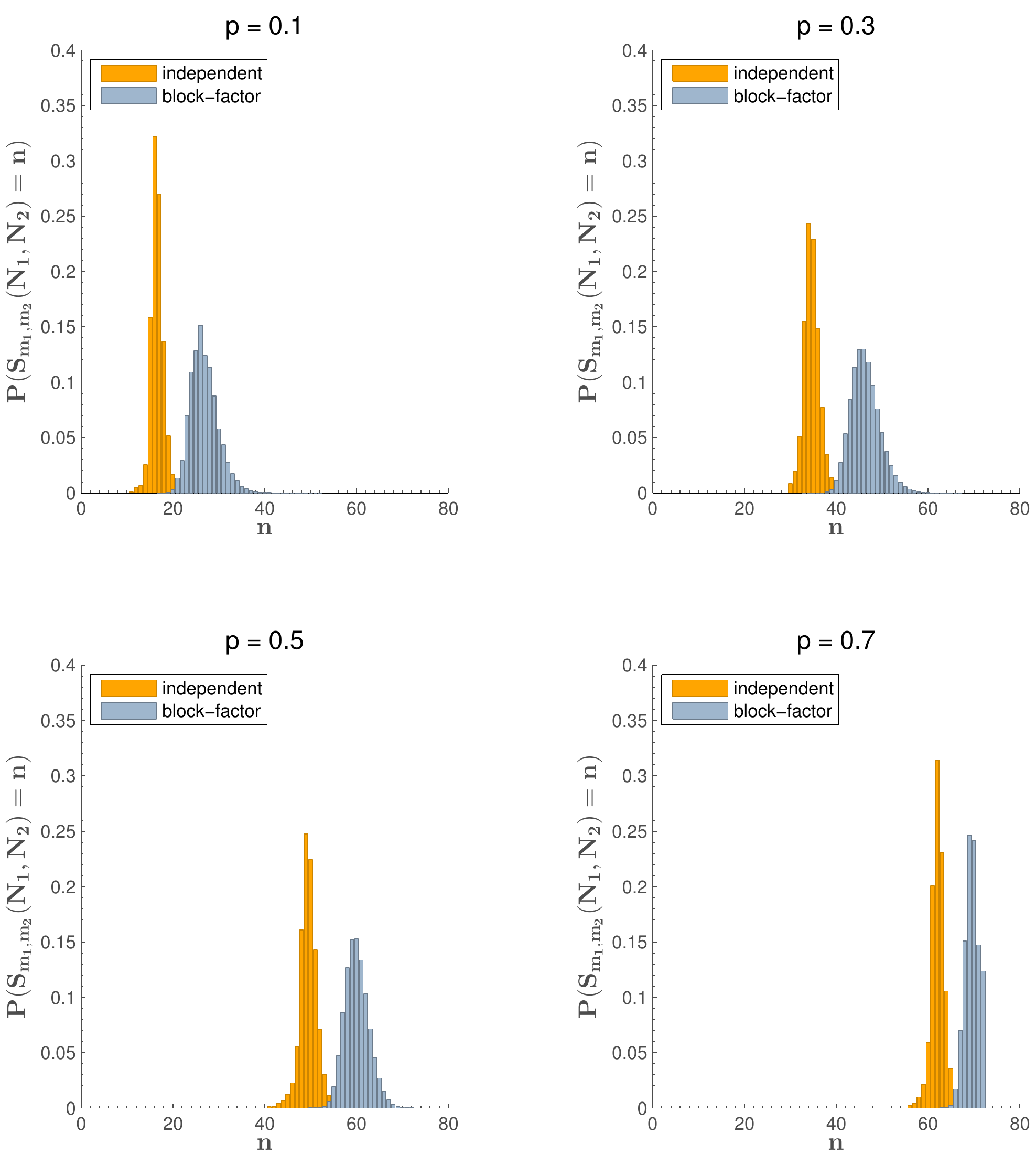}}
  \caption{Probability mass function for block−-factor and i.i.d. models}
  \label{fig7}
\end{figure}

\noindent
For all our results presented in the tables we used Monte Carlo simulations with $10^8$ iterations for the block-factor model and with $10^5$ replicas for the i.i.d. model. Notice that the contribution of the approximation error ($E_{app}$) to the total error is almost negligible in most of the cases with respect to the simulation error ($E_{sim}$). Thus, the precision of the method will depend mostly on the number of iterations ($ITER$) used to estimate $Q_{uv}$.
\noindent
The cumulative distribution function and the probability mass function for the block-factor and i.i.d. models are presented in Figure~\ref{fig6} and Figure~\ref{fig7}.

\clearpage
\subsection{Example 2: Moving Average model}\selabel{subsec4.2}
\noindent
In this example we consider the particular situation of an one dimensional scan statistics over a $MA(q)$ model. In the two dimensional block-factor model introduced in \seref{sec2} we consider $\tilde{N}_2=1$, which in particular implies that $c_2=1$ and $m_2=1$, $x_1=0$ and $x_2=q$ for $q\geq1$ a positive integer. Let $m_1\geq 2$, $\tilde{N}_{1}\geq m_1+q+1$ be positive integers and $\left\{\tilde{X}_{i}=\tilde{X}_{i,1}  \ | \  1\leq i\leq \tilde{N}_{1}\right\}$ be a sequence of i.i.d. Gaussian random variables with known mean $\mu$ and variance $\s^2$. We observe that $N_1=\tilde{N}_{1}-q$ and that for each $i\in\{1,\dots,N_1\}$ the configuration matrix becomes
\begin{equation}\label{eq4.4}
C_{(i)} = \left(\tilde{X}_{i},\tilde{X}_{i+1},\dots,\tilde{X}_{i+q}\right).
\end{equation}
Let the transformation $T:\MM_{1,q+1}(\RR)\to\RR$ be defined by
\begin{equation}\label{eq4.5}
T(x_1,\dots,x_{q+1})=a_1x_1+a_2x_2+\dots+a_{q+1}x_{q+1},
\end{equation}
where $a=(a_1,\dots,a_{q+1})\in \RR^{q+1}$ a not null vector and consider the block-factor model
\begin{equation}\label{eq4.6}
X_{i}=T\left(C_{(i)}\right)=a_1\tilde{X}_{i}+a_2\tilde{X}_{i+1}+\dots+a_{q+1}\tilde{X}_{i+q},\, \, 1\leq i\leq N_1.
\end{equation}
Clearly, the sequence $X_1,\dots,X_{N_1}$ forms a $MA(q)$ model. Notice that the moving sums $Y_{t}=Y_{t,1}$, $1\leq t\leq N_1-m_1+1$, can be expressed as
\begin{equation}\label{eq4.7}
Y_{t}=\displaystyle \sum_{i=t}^{t+m_1-1}{X_{i}}=b_1\tilde{X}_{t}+b_2\tilde{X}_{t+1}+\dots+b_{m_1+q}\tilde{X}_{t+m_1-1+q}.
\end{equation}
If, for example, $m_1\geq q$ then the coefficients $b_{1},\dots,b_{m_1+q}$ are given by
\begin{equation}\label{eq4.8}
b_k=\left \{ \begin{array}{lll}
      \displaystyle\sum_{j=1}^{k}{a_j} & \mbox{, $k\in\{1,\dots,q-1\}$}\\
      \displaystyle\sum_{j=1}^{q+1}{a_j} & \mbox{, $k\in\{q+1,\dots,m_1\}$}\\
      \displaystyle\sum_{j=k-m_1+1}^{k}{a_j} & \mbox{, $k\in\{m_1+1,\dots,m_1+q\}.$}\end{array} \right.
\end{equation}
Therefore, for each $t\in\{1,\dots,N_1-m_1+1\}$, the random variable $Y_t$ follows a normal distribution with mean $\EE\left[Y_t\right]=(b_1+\dots+b_{m+q})\mu$ and variance $Var\left[Y_t\right]=\left(b_1^2+\dots+b_{m+q}^2\right)\s^2$. The covariance matrix $\Sigma=\{Cov\left[Y_t,Y_s\right]\}$ has the entries
\begin{equation}\label{eq4.9}
Cov\left[Y_t,Y_s\right]=\left \{ \begin{array}{ll}
      \displaystyle\left(\sum_{j=1}^{m_1+q-|t-s|}{b_jb_{|t-s|+j}}\right)\s^2 & \mbox{, $|t-s|\leq m_1+q-1$}\\
      0 & \mbox{, otherwise.}\end{array} \right.
\end{equation}
Given the mean and the covariance matrix of the vector $(Y_1,\dots,Y_{N_1-m_1+1})$, one can use the importance sampling algorithm developed by \cite{Naiman2001} (see also \cite{Naiman2002} and \cite{Siegmund2007}) to estimate the distribution of the one dimensional scan statistics $S=S_{m_1}(N_1)$. Another way is to use the algorithm developed by \cite{Genz(2009)} to approximate the multivariate normal distribution. In this paper we adopt the importance sampling procedure.
\par\noindent
In order to evaluate the accuracy of the approximation developed in \seref{sec3}, we consider $q=2$, $N_1=1000$, $m_1=20$, $\tilde{X}_i\sim\MN(0,1)$ and the coefficients of the moving average model $(a_1,a_2,a_3)=(0.3,0.1,0.5)$. In Table~\ref{tab5} we present numerical results for the setting described above. In our algorithms we used $ITER_{app}=10^6$ iterations for the approximation and $ITER_{sim}=10^5$ replicas for the simulation.
\begin{table}[ht!]
    \caption{MA model: $m_1=20$, $N_1=1000$, $X_i=0.3\tilde{X}_i+0.1\tilde{X}_{i+1}+0.5\tilde{X}_{i+2}$, $ITER_{app}=10^6$, $ITER_{sim}=10^5$}
\centering
\small{
 \begin{tabular}{cccccc}
\toprule
$n$ &$Sim$  & $Approx$  & $E_{app}$  & $E_{sim}$ & $E_{total}$  \\
\midrule
11 & 0.582252 & 0.584355 & 0.011503 & 0.003653 & 0.015156  \\
12 & 0.770971 & 0.771446 & 0.002319 & 0.001691 & 0.004010  \\
13 & 0.889986 & 0.889431 & 0.000434 & 0.000733 & 0.001167  \\
14 & 0.951529 & 0.951723 & 0.000073 & 0.000297 & 0.000370  \\
15 & 0.980653 & 0.980675 & 0.000011 & 0.000113 & 0.000124  \\
16 & 0.992827 & 0.992791 & 0.000001 & 0.000040 & 0.000042  \\
17 & 0.997486 & 0.997499 & 0.000000 & 0.000013 & 0.000014  \\
18 & 0.999186 & 0.999188 & 0.000000 & 0.000004 & 0.000004  \\
19 & 0.999754 & 0.999754 & 0.000000 & 0.000001 & 0.000001  \\
20 & 0.999930 & 0.999930 & 0.000000 & 0.000000 & 0.000000  \\
\bottomrule
\end{tabular}
}
\setlength{\tabcolsep}{20em}
\label{tab5}
\end{table}
\par
\begin{figure}[ht!]
 \centerline{
  \includegraphics[width=0.8\textwidth]{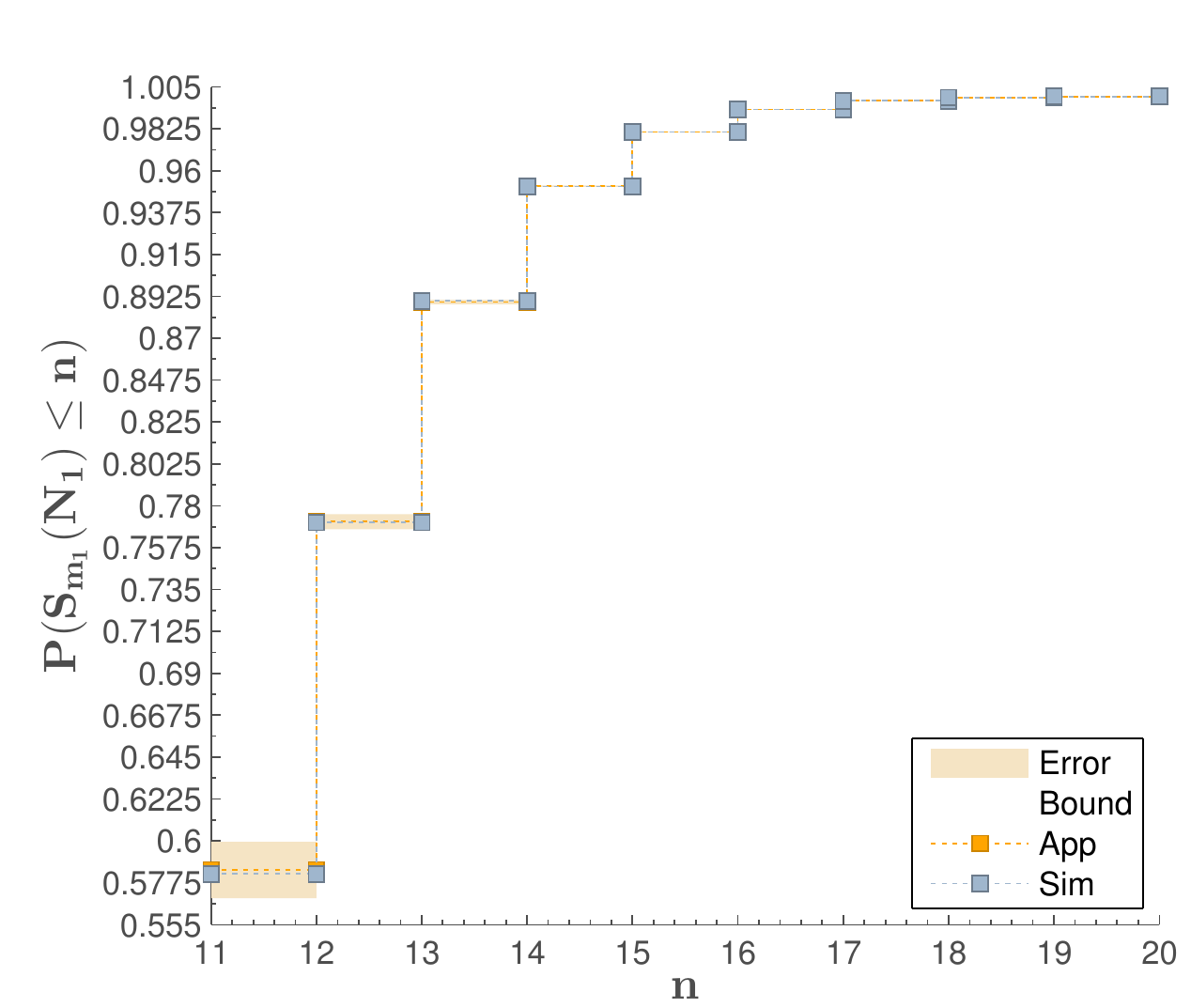}}
  \caption{Cumulative distribution function for approximation and simulation along with the corresponding error under $MA$ model}
  \label{fig8}
\end{figure}

\noindent
In Figure~\ref{fig8} we illustrate the cumulative distribution functions obtained by approximation and simulation. For the approximation we present also the corresponding lower and upper bounds (computed from the total error of the approximation process ($E_{total}$ column in Table~\ref{tab5})).

\clearpage
\section{Conclusions}\selabel{sec5}
\noindent
In this article we derived an approximation for the two dimensional discrete scan statistic generated by a block-factor type model obtained from an i.i.d. sequence. Our method provides a sharp approximation for the high order quantiles of the distribution of the scan statistics along with the corresponding error bounds. A simulation study was included to show the accuracy of our method.


\newpage
\begin{thebibliography}{99}
\bibitem[Am\u{a}rioarei(2012)]{Amarioarei}
Am\u{a}rioarei, A.: Approximation for the distribution of extremes of one dependent stationary sequences of random variables. \href{http://arxiv.org/abs/1211.5456}{arXiv:1211.5456v1}

\bibitem[Am\u{a}rioarei and Preda(2013)]{Amarioarei2013}
Am\u{a}rioarei, A. and Preda, C.: Approximation for the Distribution of Three-dimensional Discrete Scan Statistic, {\sl Methodol Comput Appl Probab}, DOI:10.1007/s11009-013-9382-3

\bibitem[Boutsikas and Koutras(2003)]{Bout2003}
Boutsikas, M. and Koutras, M. : Bounds for the distribution of two dimensional binary scan statistics, {\sl Probability in the Engineering and Information Sciences}, {\bf 17}, 509--525, 2003.

\bibitem[Boutsikas and Koutras(2000)]{Boutsikas2000}
Boutsikas, M.V., Koutras, M.: Reliability approximations for Markov chain imbeddable systems. {\sl Methodol Comput Appl Probab} {\bf 2}, 393--412, 2000.

\bibitem[Burton, Goulet and Meester(1993)]{Burton1993}
Burton, R., Goulet, M. and Meester, R.: On one-dependent processes and k-block factors. {\sl Ann. Probab.}, {\bf 21}, 2157--2168, 1993.
Processing

\bibitem[Chen and Glaz(1996)]{Glaz1996}
Chen, J. and  Glaz, J. : Two-dimensional discrete scan statistics. {\sl Statistics and Probability Letters}, {\bf 31}, 59--68, 1996.

\bibitem[Darling and Waterman(1986)]{Darling1986}
Darling, R., Waterman, M.: Extreme value distribution for the largest cube in a random lattice. {\sl SIAM J. Appl Math} {\bf 46}, 118--132, 1986.

\bibitem[Glaz, Naus and Wallenstein(2001)]{Glaz2001}
Glaz, J., Naus, J., Wallenstein, S.: Scan statistic. {\sl Springer}, 2001.

\bibitem[Glaz, Pozdnyakov and Wallenstein(2009)]{Glaz2009}
Glaz, J., Pozdnyakov, V., Wallenstein, S.: Scan statistic: Methods and Applications. {\sl Birkhauser}, 2009.

\bibitem[Genz and Bretz(2009)]{Genz(2009)}
Genz, A., Bretz, F.: Computation of Multivariate Normal and T Probabilities. {\sl Springer}, 2009.

\bibitem[Goerriero, Willett and Glaz(2009)]{Goerriero2009}
Guerriero, M., Willett, P. and Glaz, J.: Distributed target detection in a sensor network using scan statistics. {\sl IEEE Transactions on Signal}, {\sl 57}, No. 7, 2009.
Processing

\bibitem[Haiman(1999)]{Haiman}
Haiman, G.: First passage time for some stationary processes, {\sl Stochastic Processes and their Applications}, {\sl 80}, 231-248, 1999.

\bibitem[Haiman(2000)]{Haiman2}
Haiman, G.: Estimating the distributions of scan statistics with high precision, {\sl Extremes} {\bf 3:4}, 349-361, 2000.

\bibitem[Haiman (2007)]{Haiman2007}
Haiman, G.: Estimating the distribution of one-dimensional discrete scan statistics viewed as extremes of 1-dependent stationary sequences. {\sl J. Stat Plan Infer} {\bf 137}, 821--828, 2007.

\bibitem[Haiman and Preda(2002)]{HaimanPreda}
Haiman, G., Preda, C.: A new method for estimating the distribution of scan statistics for a two dimensional Poisson process. {\sl Methodology and Computing in Applied Probability} {\bf 4} 393--407, 2002.

\bibitem[Haiman and Preda(2006)]{HaimanPreda2}
Haiman, G., Preda, C.: Estimation for the distribution of two-dimensional scan statistics. {\sl Methodology and Computing in Applied Probability} {\bf 8}, 373--381, 2006.

\bibitem[Haiman and Preda(2013)]{HaimanPreda3}
Haiman, G., Preda, C.: One dimensional scan statistics generated by some dependent stationary sequences. {\sl Statistics and Probability Letters} {\bf 83}, 1457--1463, 2013.

\bibitem[Marcos and Marcos(2008)]{Marcos2008}
Marcos, R.D.L.F. and Marcos, C.D.L.F.: From star complexes to the field: open cluster families, {\sl Astrophysical Journal}, {\bf 672}, 342–351, 2008.

\bibitem[Malley, Naiman and Wilson(2002)]{Naiman2002}
Malley, J., Naiman, D., Bailey-Wilson, J.: A Comprehensive Method for Genome Scans. {\sl Human Heredity} {\bf 54}, 174-–185, 2002.

\bibitem[Naiman and Priebe(2001)]{Naiman2001}
Naiman, D., Priebe, C.: Computing Scan Statistic p Values Using Importance Sampling, with Applications to Genetics and Medical Image Analysis. {\sl J. Comp Graph Stat} {\bf 10}, 296--328, 2001.

\bibitem[Shi, Siegmund and Yakir(2001)]{Siegmund2007}
Shi, J., Siegmund, D., Yakir, B.: Importance Sampling for Estimating p Values in Linkage Analysis. {\sl Journal of the American Statistical Association} {\bf 102} (2007), 929--937.
\end{thebibliography}
\end{document}